\documentstyle[12pt]{report}
\begin{document}
\begin{center}
{\Large Phase Separation and Superconductivity in the Copper Oxide
Chain.}
\end{center}
\vspace{0.4in}
\begin{center}
C. Vermeulen$^{1}$, W. Barford$^{1}$ and E. R. Gagliano$^{2*}$
\end{center}
\begin{center}
1.Department of Physics, The University of Sheffield, Sheffield, S3 7HR, United
Kingdom.\\

2.Physics Department, University of Illinois at Urbana-Champaign,
1110 W. Green, Urbana, Il 61801, U.S.A.
 \end{center}
\vspace{1.0in}
\begin{center}
Abstract
\end{center}
The phase diagram of the copper-oxide chain as a function of density and the
nearest-neighbour Coulomb interaction, $V$, is determined.
  Phase separation takes place above a critical value of $V$ when the Cu$^{2+}
\rightarrow$ Cu$^+$ valence fluctuations dominate.  In the proximity of the
phase separation boundary the superconducting correlations are the most
divergent.  We identify the parameter regions where the Luttinger
Liquid theory applies and calculate the contours of the charge-charge
correlation exponent $K_{\rho}$. We show that
 anomalous flux quantization occurs as $K_{\rho}$ diverges.
 At $n = 1$, for $V > t$, a gap opens in the spectrum and the ground
 state is a charge-density wave.
\\
\\
$^*$Present address: Centro Atomico Bariloche, 8400 Bariloche, Argentina.

\pagebreak

\topmargin 0.0in
\footheight 0.5in
\footskip 0.5in
\oddsidemargin 0.0in
\textwidth 6.0in
\textheight 9.0in

The high temperature superconductors are difficult to understand
quantitatively because of the failure of mean field theories and
perturbation methods to give reliable results both in reduced
dimensions and in the presence of strong electronic interactions.
Under these circumstances exact calculations on finite size systems
can provide valuable information. In particular, the results of
Luttinger liquid theory enable thermodynamic correlation functions
to be deduced for one dimensional systems via exact calculations on finite
size chains. In this letter we solve exactly finite length copper-oxide chains,
up to a maximum of 12 sites, using the Lanczos procedure with the anticipation
that the copper-oxide plane will show some features in common with the
copper-oxide chain
\cite{sub1}. Our purpose is to study the phase separation and
superconducting instabilities arising from the nearest neighbour
copper-oxygen Coulomb repulsion. We identify the phase separation boundary
of
the  extended Hubbard model via a Maxwell construction.
Then we identify the region of possible superconductivity using the results
of Luttinger Liquid
theory and by the onset of anomalous flux quantization.

Our model for the copper-oxide chain consists of two atoms per unit
cell.  Neglecting the oxygen-oxygen hybridization and considering the Coulomb
interaction up to first-nearest neighbours, the copper-oxide chain is
described by the two band model Hamiltonian,
$$
H = H_t + {\Delta \over 2} \sum_{ij\sigma}{(p^{\dagger}_{j\sigma}p_{j\sigma}
    -d^{\dagger}_{i\sigma}d_{i\sigma})}
        +  U_{d} \sum_{i} d^{\dagger}_{i\uparrow}d_{i\uparrow}
        d^{\dagger}_{i\downarrow}d_{i\downarrow}
        +  U_{p} \sum_{j} p^{\dagger}_{j\uparrow}p_{j\uparrow}
        p^{\dagger}_{j\downarrow}p_{j\downarrow}\nonumber\\
$$
$$
        +  V \sum_{<ij>\sigma\sigma^\prime}
\left(d^{\dagger}_{i\sigma}d_{i\sigma}
             p^{\dagger}_{j\sigma^\prime}p_{j\sigma^\prime}\right)
         +\left( {(U_d+U_p) \over 4} + 2V \right)
         (N_{s}-N_{p}),
$$
where
$$
H_t = -t \sum_{<ij>\sigma} {(d^{\dagger}_{i\sigma}
p_{j\sigma} + h.c.) }.
\hspace{5cm} (1)
$$
\noindent $i$ and $j$ are copper and oxygen sites
respectively, $<ij>$ represents nearest neighbours and the operator
$d^{\dagger}_{i\sigma}~~(p^{\dagger}_{j\sigma})$ creates a $Cu$ $(O)$ hole with
spin $\sigma$.  $\Delta$ is the charge-transfer energy, $U_d$
$(U_p)$ is the copper (oxygen) Coulomb repulsion, $N_s$ is the number of sites,
$N_{p}$ is the number of particles, and $V$ and $t$ are the
copper-oxide Coulomb repulsion and hybridization, respectively.  The
Hamiltonian
has been defined so that it is invariant under the `particle-hole
transformation' $d^{\dagger}_{i\sigma} \to d_{i\sigma}$, $p^{\dagger}_{j\sigma}
\to p_{j\sigma}$ and $\Delta \to U_d-(\Delta+U_p)$, {\it i.e.} E($N_{p},
\Delta$) = E($2N_{s}-N_{p},U_d-(\Delta+U_p)$).

In this letter we consider parameters which are chosen to fit, as far as
possible, with the photoemission data of
high temperature superconductors. Thus $U_d\sim9t$, $\Delta\sim2-3t$ and
$U_{p}$
small, where $t\sim1.5eV$ \cite{Tun91}. Hence the region $0.5<n<1.0$
corresponds to $Cu^{2+}\rightarrow Cu^{+}$ valence fluctuations, while
(because of the `particle-hole' symmetry of equation (1)) the region
$1.5>n>1.0$
corresponds to $Cu^{2+}\rightarrow Cu^{3+}$ valence fluctuations.  As we
are interested in the role played by the nearest neighbour interaction, $V$,
this will be left as a free parameter.
\\

For most regions of the phase diagram the system is metallic, for which
 Luttinger Liquid theory can provide a valuable insight. However, for some
 parameter regions in the doping range $0.5<n<1.0$ the system is unstable with
 respect to phase separation. This occurs at a finite value of the nearest
 neighbour Coulomb repulsion, and we identify this phase boundary
 before proceeding to a discussion of the metallic regime.

  Figure 1 shows the phase separation boundary as a function of
     $V$ and hole density for $U_p=0$ and $U_p=t$. We set $U_{d}=9t$ and
$\Delta=2t$; the number of sites used was 12.  The boundary for this region is
found using the Maxwell construction.  This is given by the points
($n_{1},n_{2}$) at which a tangent touches the curve of the ground state energy
versus density.  Other authors have used the
point at which the compressibility, $\kappa$, diverges ({\it i.e.} the
point at which the ground state energy
versus density becomes concave).  These points, however, actually lie between
  $n_{1}$ and $n_{2}$ and so could be a  misleading indicator for when phase
  separation has actually occured.
Figure 1 also shows the line of $\kappa=\infty$.
  Using the Maxwell construction, the
densities of the phase separated regions are then given by $n_{1}$ and $n_{2}$,
and the fraction of the system occupied by $n_{1}$ and $n_{2}$ at any given
intermediate density $n^{\prime}$ is given by the relation
$mn_{1}+(1-m)n_{2}=n^{\prime}$, where
$m=\frac{(n_{2}-n^{\prime})}{(n_{2}-n_{1})}$.  The tendency towards charge
segregation can also be seen in the oxygen-oxygen charge correlation
function \cite{san93}, defined by
\begin{eqnarray}
C_{cdw}(k)=\frac{1}{N_{s}}\sum_{ij}(n_{i}-<n>)(n_{j}-<n>) e^{ik(R_{i}-R_{j})},
\nonumber\hspace{2cm}(2)
\end{eqnarray}
where $n_{i}=\sum_{\sigma}c^{\dagger}_{i\sigma}c_{i,\sigma}.$

The
oxygen-oxygen charge correlation function (shown in figure 2) shows a peak at
$k=\frac{\pi}{3}$ ({\it i.e.} one over the system length) for both a density of
$n=\frac{8}{12}$ and $n=\frac{10}{12}$.  This
indicates a phase separation instability in the system.
Phase separation occurs because, to avoid paying Coulomb
repulsion, holes will tend to surround themselves with empty sites. This is
because a repulsion, $V$, between a copper and a oxygen hole actually means an
attraction, $-V$, between a copper hole and a oxygen electron. The result
is that
as the doping is increased away from $n=0.5$ holes will tend to move onto
those oxygen sites which are surrounded by empty copper sites. However, where
possible holes will stay on copper sites to avoid paying the charge transfer
energy, $\Delta$. The result is that the charge density becomes segregated and
the Bloch symmetry breaks down \cite{Barford93}.

The phase separated
region is shifted as $U_p$ is increased into higher values of $V$, as can be
seen in the inset of figure 1.  However, the phase separation boundaries do
not depend strongly on $U_p$ provided that $U_p<U_d-\Delta$ is
satisfied.
\\

Having determined the region of the phase diagram for which a phase separation
instability occurs we now turn our attention to the metallic phase. In the
metallic region the long distance behaviour of the correlation functions can be
determined via the aid of Luttinger Liquid theory. Before discussing our
results
we summarise some of the important aspects of
Luttinger Liquid theory.

 Luttinger Liquids belong to the
universality class of models described by conformal field theory with central
charge, $c=1$.  This theory is useful because it gives relationships
between the correlation exponents and a few simple properties of the finite
system.  Initially it was shown that models with only one degree of freedom,
such as spinless fermions or bosons, had $c=1$.  Applying the same
ideas to the Hubbard model proved difficult since conformal quantum
field theory deals with Lorentz invariant systems only; {\it i.e.} it
requires all
gapless excitations to have the same velocity.  For the Hubbard model this is
true only at half filling, as then there is a gap for charge
excitations.

Recently it has been shown that conformal field theory can be applied to models
with more than one gapless excitation if the excitations are decoupled in the
low energy regime.  In the low energy regime this is true of the Hubbard model
\cite{Frahm90}.  In such a case each degree of freedom
has an associated operator algebra, which is again characterized by a central
charge $c=1$. Provided that there are no phase changes as the interactions are
 increased it is reasonable to assume that Luttinger Liquid theory applies to
 lattice quantum models in the strong coupling regime. In fact, this assumption
 can be tested {\it a posteriori} by showing that the predictions of
Luttinger Liquid theory are internally consistent. This is the assumption
adopted in this paper
 for the extended Hubbard model.

At long distances the charge-charge and spin-spin correlation
functions are given by \cite{schulz90},
$$
<n(x)n(0)> = {K_{\rho} \over {(\pi x)^2}} +
            A_{1} {cos~{(2k_{f}x)} \over {x^{(1+K_{\rho})}}}
            ln^{-3/2}(x)
            +A_{2} {cos~{(4k_{f}x)} \over {x^{4K_{\rho}}}}
$$
and
$$
<S_{z}(x)S_{z}(0)> = {1 \over {(\pi x)^2}} +
            B_{1} {cos~{(2k_{f}x)} \over {x^{(1+K_{\rho})}}}
            ln^{1/2}(x),
$$
respectively.  The superconducting correlations are similar to
the spin-spin correlator with
$K_{\rho}$ being replaced by $1/K_{\rho}$.  Hence, $K_{\rho}$ equals 1 for
free fermions with spin.  $K_{\rho} < 1$ means that the charge
and spin correlations dominate and $K_{\rho} > 1$ implies that the
superconducting correlations are the most divergent.

    Conformal field theory relates bulk quantities such as the
compressibility, $\kappa$, and the Drude weight, $D_c$, with the charge-charge
correlation exponent.  Where the system behaves as a Luttinger Liquid,
the exponent $K_{\rho}$ is related to the compressibility by $K_{\rho} = {\pi
\over 2} {\kappa v_{c} n^2}$ and to the Drude weight by
$K_{\rho}=D_{c}\pi/v_{c}$ (where $v_{c}$
is the charge velocity).  When analytic solutions are not available these bulk
quantities \cite{ogata91,lin93} can be evaluated using exact diagonalization
techniques \cite{gagliano86}.  The discrete compressibility is given by
${N_{p}^2}\kappa = 4/ [e(N_{p}+2)+e(N_{p}-2)-2e(N_{p})]$, where $e(N_{p})$ is
the ground state energy per site.
To calculate the Drude weight the chain is threaded
with a dimensionless constant flux, $\phi$.
This is done mathematically by the operator transformation, $c^{\dagger}_
{k\sigma}\rightarrow c^{\dagger}_{k\sigma}e^{\frac{-ik\phi}{N_{s}}}$.
  The finite-size equivalent of the
Drude weight is then $D_{c}= [e(\phi_{0}+\delta \phi)-e(\phi_{0})] /\delta
\phi^{2}$,
where $\delta \phi$ is an infinitesimal flux $(\sim 0.01)$ and $\phi_{0}$ is a
background flux which removes accidental degeneracies of the ground state and
guarantees a $zero-current$ ground state.  For example,
$\phi_0$ is $0$ ({\it periodic boundary conditions}) for $N_{p}=4m+2$ and
${\pi \over {N_s}}$  ({\it anti-periodic boundary conditions}) for $N_{p}=4m$
with $m$ an integer \cite{comment0}.

The charge velocity, as well as the spin velocity, can also be obtained from
finite cluster
calculations of the response functions \cite{gagliano87}. In fact, they are
related to the lowest frequency pole $\omega_{l}$ of the charge-charge and
spin-spin dynamical correlation functions by $v_{l}={\omega_{l} \over
{q_{min}}}$, where $q_{min}={2\pi \over {N_{u}}}$,  $N_{u}$ is the number
of unit
cells and $l$ refers to both charge and spin degrees of freedom
\cite{comment1}.
One can therefore directly determine both velocities without adiabatically
following the relevant low-energy excitations, as has recently been done for
the
$t-J$ model \cite{ogata91}.

The phase diagram, figure 1, shows the contour of $K_{\rho}=1$, and the
region for which $K_{\rho}$ exceeds $1$. The expression used to calculate this
contour was $K_{\rho}={\pi}n\sqrt{ {D_{c}\kappa \over 2} }$. Notice that for
low doping $K_{\rho}$ exceeds $1$ within the phase separation region given
by the Maxwell construction, and as doping is increased the
curve $K_{\rho}=1$ moves out indicating
  that the superconducting correlations begin to dominate well within the
conducting region. The points on the contour corresponding to the densities
$n=\frac{8}{12}$ and $n=\frac{10}{12}$ are found from the
12 site chain, the point at $n=\frac{8}{10}$ is found using a 10 site chain
and  the point at $n=\frac{6}{8}$ is
found using an 8 site chain. At high densities the point of
  $K_{\rho}$ exceeding $1$ is
associated with the appearance of anomalous flux quantization.
 Anomalous flux quantization is the condition that the ground state energy
is periodic in $\phi$, with period $\pi$,
 ({\it i.e.} the flux is quantised in units of $\frac{hc}{2e}$).
   For the attractive Hubbard model, which is known to exhibit singlet
superconductivity, the ground state energy satisfies
\begin{eqnarray}
E_{0}(\phi+\pi)-E_{0}(\phi)=\frac{\Lambda}{2}
\left(\frac{N_{s}}{\zeta}\right)^{1/2}
exp\left(\frac{-N_{s}}{\zeta}\right), \nonumber \hspace{2cm}(3)
\end{eqnarray}
\noindent where $\zeta$ is a
length associated with a gap of spin excitations and $\Lambda$ depends on
details of the model \cite{staff91}. As well as exhibiting anomalous flux
quantization, the superconducting state is defined by the Drude weight
(or superfluid density) remaining finite in the thermodynamic limit.
For $n=\frac{10}{12}$  and $\frac{8}{10}$ anomalous flux quantization
appears almost exactly at the point that $K_{\rho}$ exceeds $1$,
the result being slightly better for $n=\frac{10}{12}$.
For lower densities this relationship is not so clear.

We have also performed finite size scaling \cite{sub93} using
  a 6 and 12 site chain at $n=\frac{2}{3}$.  Using these two chains we
are able to calculate the parameter $\zeta$ of equation (3) and the ratio
$\frac{D_{c}(Ns=12)}{D_{c}(Ns=6)}$ for different values of $V$.  We are
then able
to determine whether this is in the asymptotic region of equation (3) ({\it
i.e.} the region of relatively large $N_{s}$ and small $\zeta$).  A small value
of the ratio of the Drude weights in the asymptotic region will indicate that
$D_{c}=0$ in the thermodynamic limit, and that the true ground state is
non-conducting.  Anomalous flux quantization, without conductivity, is the
signal for a pairing but non-superconducting state, such as a phase separated
state or a charge density wave.  When $V=4t$
(a point well within the phase separation region),
 $\zeta\sim2$ so this is in the asymptotic region, but the ratio
$\frac{D_{c}(Ns=12)}{D_{c}(Ns=6)}\sim0.1$.  Such a large reduction in Drude
 weight with such a small $\zeta$ indicates an insulating state.  Anomalous
flux
quantization first occurs when $V\sim2.7t$ for a 12 site chain at this
doping.  This indicates that at $V=4t$ there is a definite phase separation
instability.

For all other values of doping, namely $n<0.5$ and $n>1$, the
effect of nearest neighbour repulsion is to suppress the superconducting
correlations.  For the densities $n<0.5$ and $n>1.5$, the phase diagram
rather resembles the corresponding phase diagram of the $t-U-V$
model \cite{lin93}.

Finally, we consider the effect of the nearest neighbour repulsion at half
filling, $n=1$. At this filling the effect of increasing $V$ is to force the
system into an insulating phase.  This can be seen from the graph of the Drude
weight, the inset of figure 3, which vanishes at $V\sim1.6t$.  The driving
force for this change is illustrated by the charge correlation function shown
in
figure 3. It shows that as $V$ is increased there is a sharp rise in the
correlation function at $k=\pi$ (a wavelength of {\it one} unit cell).  This
 corresponds to charge being pushed from copper sites onto oxygen sites,
resulting in an insulating charge density wave.
\\

 In conclusion,
we have established the phase diagram of the copper-oxide chain using finite
cluster calculations.  In the region in which $Cu^{2+}\rightarrow Cu^{+}$
charge
fluctuations dominate ({\it i.e.} $\Delta << U_{d}$) we find that the
coupling to the
nearest neighbour repulsion leads to a region of phase separation for a
sufficiently large interaction (typically $V \sim \Delta$).  In the proximity
of
the phase separation region and large enough density superconducting
correlations are the most divergent.  From our results we find that the curve
$K_\rho=1$ crosses the phase separation boundary at $n\sim0.72$. At the
densities
$n=\frac{8}{10}$ and $\frac{10}{12}$ the divergence of $K_\rho$ is accompanied
by anomalous flux quantization.
 The contour $K_\rho =1$ has been calculated near the metal-insulator
transition
 ($n=0.5$) with only a small number of doped holes. The trend for the
superconducting region to disappear as $n\rightarrow0.5$ may be a consequence
of this fact.  Close to the phase separation boundary the superconducting
correlations are precursory to phase separation. At weaker coupling the role
played by charge transfer excitons \cite{Vermeulen94} is still to be
understood.
For all other parameter regions superconducting correlations are suppressed.
At
$n=1$ the model shows a transition from a metallic state to a insulating charge
density wave.
\vspace {1 cm}

\begin{center}
Acknowledgements\\
\end{center}

We thank the SERC (United Kingdom) for the provision of a Visiting Research
Fellowship (ref. GR/H33091). W.B. also acknowledges support from the
SERC (ref. GR/F75445) and a grant from the University of Sheffield research
fund,
while E.G. acknowledges support from DMR89-20538/24. C.J.V. is supported by a
University of Sheffield scholarship.  We thank
B. Alascio, A. Aligia, S.  Bacci, C. Balseiro, E. Fradkin, G. A. Gehring and
R. M. Martin for useful conversations.  The
computer calculations were performed at the National Center for
Supercomputing Applications, Urbana, Illinois and at the University of
Sheffield.

 \pagebreak

\begin{figure}
{\bf Figure 1:} The phase diagram of the copper-oxide chain as a function of
$V$
and hole density for $U_{d} = 9t$, $U_{p}=0$ and $\Delta =2t$, showing the
phase separation boundary, the contour of $K_{\rho}$=1 and the contour of
infinite compressibility.  There is also shown the onset of anomalous flux
quantization for $n=\frac{10}{12}$ and $\frac{8}{10}$. The inset shows the
phase
separation boundary for $U_p=0$ and $t$.
\\
\\
\\
\\
{\bf Figure 2:} The
oxygen-oxygen charge correlation function for $n=\frac{10}{12}$ and
$\frac{8}{12}$. This shows a peak at $k=\pi/3$ for $V=2t$, indicating a phase
separated state ($U_d=9t$, $U_p=0$ and $\Delta=2t$)
\\
\\
\\
\\
{\bf Figure 3:}
The Drude weight as a function of hole density for $U_d=9t$, $U_p=0$ and
 $\Delta=2t$, showing the opening of the gap at $n=1$.  The copper-oxygen
charge-charge correlation function for a 12 site chain at $n=1$ is shown for
$V=0.5t$ and $2t$.  The density of electrons on the oxygen atoms rises from
$n=1.265$ to $n=1.797$ in this range.  \end{figure}


\begin{thebibliography}{99}
\bibitem{sub1}
These studies are also interesting in their own right, however, because
copper-oxide chains do exist in some of the high temperature superconductors,
for example YBa$_2$Cu$_3$O$_7$ (in the form of copper-oxide diamonds).
Furthermore, the one dimensional halogen-bridged metal complexes are
also charge transfer systems which exhibit many of the
fascinating features of polyacetylene, for example charge density waves, spin
density waves and solitons.  To model true one dimensional systems,
however, electron-phonon couplings should be included to account for
the Peierls transition. This will not be done in this paper.
\bibitem{comment0} For a two-band spinless
Fermion model, antiperiodic boundary conditions are required at any density in
order to have a zero current ground state.
\bibitem{comment1} Since the charge operator does not conserve total spin, the
peak associated with $v_s$, appears also in the charge-charge response
function.
Hence, the charge velocity is given by the low energy peak that does not appear
simultaneously in both dynamical correlations.
\end{thebibliography}
\end{document}